\documentclass[twocolumn,showpacs,preprintnumbers,nofootinbib,amsmath,amssymb]{revtex4}

\usepackage{euscript,amssymb,amsmath}

\usepackage{graphicx}
\usepackage{epsfig}

\newcommand{\be}[1]{\begin{equation}\label{#1}}
\newcommand{\ee}{\end{equation}}
\newcommand{\ba}[1]{\begin{eqnarray}\label{#1}}
\newcommand{\ea}{\end{eqnarray}}
\newcommand{\rf}[1]{(\ref{#1})}
\newcommand{\nn}{\nonumber}

\begin{document}

\title{Weak-field limit of Kaluza-Klein models with spherical compactification: \\experimental constraints}

\author{Alexey Chopovsky}\email{alexey.chopovsky@gmail.com}  \author{Maxim Eingorn}\email{maxim.eingorn@gmail.com}  \author{Alexander Zhuk}\email{ai_zhuk2@rambler.ru}

\affiliation{Astronomical Observatory and Department of Theoretical Physics, Odessa National University, Street Dvoryanskaya 2, Odessa 65082, Ukraine\\}

\begin{abstract} We investigate the classical gravitational tests for the six-dimensional Kaluza-Klein model with spherical (of a radius $a$) compactification of the
internal space. The model contains also a bare multidimensional cosmological constant $\Lambda_6$. The matter, which corresponds to this ansatz, can be simulated by a
perfect fluid with the vacuum equation of state in the external space and an arbitrary equation of state with the parameter $\omega_1$ in the internal space. For
example, $\omega_1=1$ and $\omega_1=2$ correspond to the monopole two-forms and the Casimir effect, respectively. In the particular case $\Lambda_6=0$, the parameter
$\omega_1$ is also absent: $\omega_1=0$. In the weak-field approximation, we perturb the background ansatz by a point-like mass. We demonstrate that in the case
$\omega_1>0$ the perturbed metric coefficients have the Yukawa type corrections with respect to the usual Newtonian gravitational potential. The inverse square law
experiments restrict the parameters of the model: $a/\sqrt{\omega_1}\lesssim 6\times10^{-3}\ {\mbox{cm}}$. Therefore, in the Solar system the parameterized
post-Newtonian parameter $\gamma$ is equal to 1 with very high accuracy. Thus, our model satisfies the gravitational experiments (the deflection of light and the time
delay of radar echoes) at the same level of accuracy as General Relativity. We demonstrate also that our background matter provides the stable compactification of the
internal space in the case $\omega_1>0$. However, if $\omega_1=0$, then the parameterized post-Newtonian parameter $\gamma=1/3$, which strongly contradicts the
observations.
\end{abstract}

\pacs{04.25.Nx, 04.50.Cd, 04.80.Cc, 11.25.Mj}

\maketitle

\vspace{.5cm}

%%%%%%%%%%%%%%%%%%%%%%%%%%%%%%%%%%%%%%%%%%%%%%%%%%%%%%%%%%%%%%%%%

\section{\label{sec:1}Introduction}

Any physical theory is correct until it does not conflict with the experimental data. Obviously, the Kaluza-Klein model is no exception to this rule.
%This model is a modification of General Relativity to the number of dimensions of space-time greater than four.
There is a number of well-known gravitational experiments in the Solar system, e.g., the deflection of light, the perihelion shift and the time delay of radar echoes
(the Shapiro time-delay effect). In the weak-field limit, all these effects can be expressed via parameterized post-Newtonian (PPN) parameters $\beta$ and $\gamma$
\cite{Will,Straumann}. These parameters take different values in different gravitational theories. There are strict experimental restrictions on these parameters
\cite{Bertotti,Will2,JKh,CFPS}. The tightest constraint on $\gamma$ comes from the Shapiro time-delay experiment using the Cassini spacecraft: $\gamma -1 = (2.1\pm
2.3)\times 10^{-5}$. General Relativity is in good agreement with all gravitational experiments \cite{Landau}. Here, the PPN parameters $\beta=1$ and $\gamma=1$. The
Kaluza-Klein model should also be tested by the above-mentioned experiments.

In our previous papers \cite{EZ3,EZ4,EZ5} we have investigated this problem in the case of toroidal compactification of internal spaces. We have supposed that in the
absence of gravitating masses the metrics is a flat one. Gravitating compact objects (point-like masses or extended massive bodies) perturb this metrics, and we have
considered these perturbations in the weak-field approximation. First, we have shown that in the case of three-dimensional external/our space and dust-like equations of
state \cite{footnote} in the external and internal spaces, the PPN parameter $\gamma = 1/(D-2)$, where $D$ is a total number of spatial dimensions. Obviously, $D=3$
(i.e. the General Relativity case) is the only value which does not contradict the observations \cite{EZ3}. Second, in the papers \cite{EZ4,EZ5}, we have investigated
the exact soliton solutions. In these solutions a gravitating source is uniformly smeared over the internal space and the non-relativistic gravitational potential
exactly coincides with the Newtonian one. Here, we have found a class of solutions which are indistinguishable from General Relativity. We have called such solutions
latent solitons. Black strings and black branes belong to this class. They have the dust-like equation of state $p_0=0$ in the external space and the relativistic
equation of state $p_1=-\varepsilon/2$ in the internal space. It is known (see \cite{EZ5,Zhuk}) that in the case of the three-dimensional external space with a dust-like
perfect fluid, this combination of equations of state in the external and internal spaces does not spoil the internal space stabilization. Moreover, we have shown also
that the number $d_0=3$ of the external dimensions is unique. Therefore, there is no problem for black strings and black branes to satisfy the gravitational experiments
in the Solar system at the same level of accuracy as General Relativity. However, the main problem with the black strings/branes is to find a physically reasonable
mechanism which can explain how the ordinary particles forming the astrophysical objects can acquire rather specific equations of state $p_i=-\varepsilon/2$ (tension!)
in the internal spaces. Thus, in the case of toroidal compactification, on the one hand we arrive at the contradiction with the experimental data for the physically
reasonable gravitating source in the form of a point-like mass, on the other hand we have no problem with the experiments for black strings/branes but arrive at very
strange equation of state in the internal spaces. How common is this problem for the Kaluza-Klein models?

To understand it, in the present paper we investigate a model with spherical compactification of the internal space. Therefore, in contrast to the previous case the
background metrics is not flat but has a topology $\mathbb{R}\times\mathbb{R}^3\times S^2$. To make the internal space curved, we must introduce a background matter. We
show that this matter can be simulated by a perfect fluid with the vacuum equation of state in the external space and an arbitrary equation of state with the parameter
$\omega_1$ in the internal space. Our model contains also a bare multidimensional cosmological constant $\Lambda_6$. If $\Lambda_6$ is absent, then the parameter
$\omega_1$ is also equal to zero, i.e. the perfect fluid has the dust-like equation of state in the internal space. We perturb this background by a point-like mass and
calculate the perturbed metric coefficients in the weak-field approximation up to the order $1/c^2$. In the case $\omega_1 >0$, these metric coefficients acquire the
Yukawa correction terms with respect to the usual Newtonian gravitational potential. The Yukawa interaction is characterized by its mass which is proportional to
$\sqrt{\omega_1}$. The terrestrial inverse square law experiments \cite{new} restrict such corrections and provide strong bounds on parameters of the model, e.g., on a
radius of the internal two-sphere. We show that this radius is in many orders of magnitude less than the radius of the Sun. Obviously, in the Solar system we can drop
the Yukawa correction terms with very high accuracy, and the parameterized post-Newtonian parameter $\gamma$ is equal to 1 similar to General Relativity. Therefore, our
model satisfies the gravitational experiments (the deflection of light and the time delay of radar echoes) at the same level of accuracy as General Relativity. On the
other hand, in the case $\Lambda_6=0$, $\omega_1=0$ (i.e. the Yukawa mass is equal to zero), the PPN parameter $\gamma=1/3$ which exactly coincides with the formula
$\gamma =1/(D-2)$ for $D=5$. Obviously, this value contradicts the observations. It is worth noting that $\omega_1>0$ is the sufficient condition of the internal space
stabilization in our model. Therefore, we demonstrate that the main problem of the case $\Lambda_6=0$ (as well as of the models in the papers \cite{EZ3,EZ4,EZ5}) is that
the internal space is not stabilized. We show that the inclusion of the background matter which stabilizes the internal two-sphere can solve the problem.

The paper is organized as follows. In section II we define the background metrics and matter for the Kaluza-Klein model with flat external space-time and spherical
compactification of the internal space. We also include a bare six-dimensional cosmological constant $\Lambda_6$. We perturb this background by a point-like mass and
calculate the corresponding perturbed metric coefficients. Then, we
%define the parameter space which can provide the agreement with the observations.
define the conditions which provide the agreement with the observations. One of these conditions is the positivity of the equation of state parameter $\omega_1$ in the
internal space. In appendixes A and B we present formulas for the components of the Ricci tensor   and, with the help of them, investigate the relations between the
perturbed metric coefficients. In appendix C we prove that the background matter which satisfies the condition $\omega_1>0$ stabilizes the internal two-sphere. The main
results are summarized and discussed in section III.

\section{\label{sec:2}Background solution and perturbations}

To start with, let us consider a factorizable six-dimensional static background metrics
%%%%%%%
\be{c2.1}
 ds^2=c^2 dt^2-dx^2-dy^2-dz^2-a^2(d\xi^2+\sin^2\xi d\eta^2)\, ,
\ee
%%%%%%%
which is defined on a product manifold $M = M_4\times M_2$. $M_4$ describes external four-dimensional flat space-time and $M_2$ corresponds to the two-dimensional
internal space which is a sphere with the radius (the internal space scale factor) $a$. We also include in the model a bare multidimensional
cosmological constant $\Lambda_6$. As we shall see below, we need such term to stabilize the internal space. Therefore, the Einstein equation reads
%%%%%%%%
\be{c2.2} \kappa T_{ik} = R_{ik}-\frac{1}{2} R g_{ik}-\kappa\Lambda_6g_{ik}\, , \ee
%%%%%%
where $\kappa \equiv 2S_5\tilde G_{6}/c^4$. Here, $S_5=2\pi^{5/2}/\Gamma (5/2)=8\pi^2/3$ is the total solid angle (the surface area of the four-dimensional sphere of a
unit radius) and $\tilde G_{6}$ is the gravitational constant in the six-dimensional space-time.

According to appendix \ref{sec:A}, the only nonzero components of the Ricci tensor for the metrics \rf{c2.1} are $R_{44}=1$ and $R_{55}=\sin^2 \xi$, and the scalar
curvature is $R=-2/ a^2$. Therefore, the energy-momentum tensor which corresponds to the background metrics \rf{c2.1} is
%%%%%%
\be{c2.3}
T_{ik}=\left\{
\begin{array}{cc}
\left( 1/\left(\kappa a^2\right)-\Lambda_6 \right) g_{ik} & \mbox{for   } \, i,k=0,...,3;\\
\\
-\Lambda_6 g_{ik} & \mbox{for   } \, i, k=4,5.
\end{array} \right . \quad
\ee
%%%%%%%%
This expression can be written in the form of the energy-momentum tensor of a perfect fluid:
%%%%%%%
\be{c2.4}
T^i_k = \mbox{diag}\left(\bar{\varepsilon}, -\bar{p}_0,-\bar{p}_0,-\bar{p}_0,-\bar{p}_1,-\bar{p}_1\right)\, ,
\ee
%%%%%%%
where the energy density and pressures in the external and internal spaces are respectively
%%%%%%%%%
\be{c2.5}
\bar{\varepsilon} \equiv \frac{1}{\kappa a^2}-\Lambda_6\, ,\quad \bar{p}_0 \equiv -\left(\frac{1}{\kappa a^2}-\Lambda_6\right)\, ,\quad
\bar{p}_1\equiv \Lambda_6\, .
\ee
%%%%%%%
The upper bar denotes the background values. Therefore, the equation of state in the external space reads
%%%%%%%
\be{c2.6}
\bar{p}_0=\omega_0 \bar \varepsilon\, , \quad \omega_0 =-1,
\ee
%%%%%
that is we have the vacuum-like equation of state in the external space, but the equation of state in the internal space is not fixed:
%%%%%%%
\ba{c2.7} &{}&\bar{p}_1=\omega_1\bar \varepsilon \quad \Rightarrow \nn\\
&{}& \omega_1 = \frac{\Lambda_6}{1/\left(\kappa a^2\right)-\Lambda_6}\, \Leftrightarrow \, \Lambda_6 = \frac {\omega_1}{\omega_1+1}\; \frac{1}{\kappa a^2}\, , \ea
%%%%%%
i.e. $\omega_1$ is arbitrary. The case $\omega_1=0$ automatically results in $\Lambda_6=0$. Choosing different values of $\omega_1$ (with
fixed $\omega_0 = -1$), we can simulate different forms of matter. For example, $\omega_1=1$ and $\omega_1=2$ correspond to the monopole form-fields (the Freund-Rubin
scheme of compactification) and the Casimir effect, respectively (see appendix \ref{sec:C} and \cite{FR,exci,Zhuk}).

Now, we perturb our background ansatz by a static point-like massive source with non-relativistic rest mass density $\rho$. We suppose that the matter source is
uniformly smeared over the internal space \cite{footnote 2}. Hence, multidimensional $\rho$ and three-dimensional $\rho_3$ rest mass densities are connected as follows:
$\rho=\rho_3({\bf r}_3)/\left(4\pi a^2\right)$ \cite{EZ3,EZ2}. In the case of a point-like mass $m$, $\rho_3(r_3)=m\delta ({\bf r}_3)$, where $r_3=|{\bf
r}_3|=\sqrt{x^2+y^2+z^2}$. In the non-relativistic approximation the only nonzero component of the energy-momentum tensor of the point-like mass is $\hat{T}^0_{0}\approx
\rho c^2$ and up to linear in perturbations terms $\hat{T}_{00}\approx \rho c^2$. Concerning the energy-momentum tensor of the background matter, we suppose that
perturbation does not change the equations of state in the external and internal spaces, i.e. $\omega_0$ and $\omega_1$ are constants. For example, if we had a monopole
form-fields ($\omega_0=-1,\,\omega_1=1$) before the perturbation, the same type of matter we shall have after the perturbation. Therefore, the energy-momentum tensor of
the perturbed background is
%%%%%%
\be{c2.8}
\tilde T_{ik}\approx\left\{
   \begin{array}{cc}
   \left(\bar\varepsilon +\varepsilon^1\right) g_{ik}, & i,k=0,...,3;\\
   \\
   -\omega_1 \left(\bar\varepsilon +\varepsilon^1\right) g_{ik}, & i, k=4,5\, ,
   \end{array}
\right.
\ee
%%%%%%%%%
where the correction $\varepsilon^1$ is of the same order of magnitude as the perturbation $\rho c^2$. The trace of \rf{c2.8} is $T\approx2(2-\omega_1)(\bar\varepsilon
+\varepsilon^1)$.

We suppose that the perturbed metrics preserves its diagonal form. Obviously, the off-diagonal coefficients $g_{0\alpha}, \, \alpha =1,\ldots ,5$, are absent for the
static metrics. It is also clear that in the case of uniformly smeared (over the internal space) perturbation, all metric coefficients depend only on $x,y,z$ (see, e.g.,
\cite{EZ2}), and the metric structure of the internal space does not change, i.e. $F=E\sin ^2\xi$. It is not difficult to show that in this case the spatial part of the
external metrics can be diagonalized by coordinate transformations. Moreover, if we additionally assume the spherical symmetry of the perturbation with respect to the
external space, then all metric coefficients depend on $r_3$ and $B(r_3)=C(r_3)=D(r_3)$. Therefore, the perturbed metrics reads
%%%%%%%
\be{c2.9} ds^2=Ac^2dt^2+Bdx^2+Cdy^2+Ddz^2+Ed\xi^2+Fd\eta^2 \ee
%%%%%%%
with
%%%%%%%
\ba{c2.10}
&{}& A \approx 1+A^{1}({r}_3),\quad B \approx -1+B^{1}({r}_3)\, ,\nn\\
&{}& C \approx -1+C^{1}({r}_3),\quad D \approx -1+D^{1}({r}_3)\, ,\nn\\
&{}& E \approx -a^2+E^{1}({r}_3),\quad F \approx -a^2\sin^2\xi+F^{1}({r}_3) \, ,\nn\\
&{}& \ea
%%%%%%%
where we take into account the spherical symmetry of the perturbation with respect to the external space. All perturbed metric coefficients $A^1,B^1,C^1,D^1,E^1$ and
$F^1$ are of the order of $\varepsilon^1$. To find these coefficients we should solve the Einstein equation
%%%%%%%
\be{c2.11}
R_{ik}=\kappa\left( T_{ik}-\frac{1}{4}Tg_{ik} -\frac12 \Lambda_6 g_{ik}\right)\, ,
\ee
%%%%%%%
where the energy-momentum tensor $T_{ik}$ is the sum of the perturbed background $\tilde T_{ik}$ \rf{c2.8} and the energy-momentum tensor of the perturbation $\hat
T_{ik}$. Then,
%it can be easily seen that instead of equations (16), (17) and (18) in \cite{ChEZ1}
we get the system
%%%%%
\ba{c2.12} &{}& \triangle_3 A^1=\kappa\omega_1\varepsilon^1+\frac{3}{2}\kappa\rho c^2\, ,\nn\\
&{}& \triangle_3 B^1=\triangle_3 C^1=\triangle_3 D^1=-\kappa\omega_1\varepsilon^1+
\frac{1}{2}\kappa\rho c^2\, ,\\
\label{c2.13} &{}& \triangle_3 E^1=(2+\omega_1)\kappa a^2\varepsilon^1-\frac{2}{a^2}E^1+\frac{1}{2}\kappa\rho c^2a^2\, , \ea
%%%%%%
where $\triangle_3$ is the three-dimensional Laplace operator. Eqs. \rf{c2.12} show that $B^1=C^1=D^1$ and the relation $B^1=(1/3)A^1$ takes place only in the particular
case  $\omega_1=0$ (the case of \cite{ChEZ1}).
%Therefore, $B^1\ne (1/3)A^1$ in general case $\omega_1\ne 0$.
%According to appendix B case 1 ("Smeared extra dimensions") of \cite{ChEZ1}, the perturbed metric coefficients should satisfy the %following two conditions:
%%%%%%
%\be{c2.14}
%-A^1+B^1+\frac{2}{a^2}E^1=0,\quad F^1=E^1\sin^2\xi\, .
%\ee
%%%%%%
With the help of Eqs. \rf{b9} and \rf{c2.12} we obtain
%%%%%
\be{c2.15}
\triangle_3 E^1=\frac{a^2}{2}\left(\triangle_3 A^1-\triangle_3 B^1\right)=\frac{a^2}{2}(2\kappa\omega_1\varepsilon^1+\kappa\rho c^2)\, .
\ee
%%%%%%
%where we take into account \rf{c2.12}.
The comparison of \rf{c2.13} and \rf{c2.15} yields
%%%%%%%
\be{c2.16}
\kappa \varepsilon^1 = \frac{E^1}{a^4}\, .
\ee
%%%%%%%
The substitution of this relation back into \rf{c2.13} gives
%%%%%%%
\be{c2.17}
\frac{\omega_1}{a^2}E^1=\triangle_3 E^1-\frac{1}{2}\kappa\rho c^2 a^2\, .
\ee
%%%%%%%
Then, taking also into account \rf{c2.16}, we can rewrite Eqs. \rf{c2.12} in the form
%%%%%%
\be{c2.18}
\triangle_3\left(A^1-\frac{E^1}{a^2}\right)=\kappa\rho c^2,\quad\triangle_3\left(B^1+\frac{E^1}{a^2}\right)=\kappa\rho c^2\, .
\ee
%%%%%%
In the case of smeared extra dimensions the rest mass density is $\rho = \left(m/\left(4\pi a^2\right)\right)\delta ({\bf r}_3)$. Hence, the Eq. \rf{c2.17} can be
rewritten as follows
%%%%%
\be{c2.19} \triangle_3 E^1 -\lambda^{-2} E^1 = -\nu \delta ({\bf r}_3)\, , \ee
%%%%%
where parameters $\lambda^2 \equiv a^2/\omega_1$ and $\nu \equiv -a^2 4\pi G_N m/c^2$. We also introduce the Newton gravitational constant via the relation
%%%%%
\be{c2.20}
4\pi G_N = \frac{S_5 \tilde G_6}{4\pi a^2}\, ,
\ee
%%%%%
which exactly coincides with the formula (58) in \cite{EZ4} where the volume of the internal space $V_2=4\pi a^2$. It is well known that to get the solution of \rf{c2.19}
with the boundary condition $E^1 \to 0$ for $r_3 \to +\infty$ the parameter $\lambda^2$ should be positive, i.e. the equation of state parameter $\omega_1$ should
satisfy the condition
%%%%%%
\be{c2.21}
\omega_1> 0\, .
\ee
%%%%%%
Additionally, we can conclude from \rf{c2.7} that the bare six-dimensional cosmological constant is also positive: $\Lambda_6>0$.
In the case of positive $\omega_1$, the solution of \rf{c2.19} reads
%%%%%%%
\be{c2.22}
E^1=\frac{\nu}{4\pi r_3}e^{-r_3/\lambda}=a^2 \frac{\varphi_N}{c^2}e^{-r_3/\lambda}\, ,
\ee
%%%%%%%
where the Newtonian potential $\varphi_N=-G_N m/r_3$. Now, we can easily get the solutions of Eqs. \rf{c2.18}:
%%%%%%%
\ba{c2.23}
A^1&=&\frac{2\varphi_N}{c^2}+\frac{E^1}{a^2}=\frac{2\varphi_N}{c^2}\left[1+\frac{1}{2}\exp\left(-r_3/\lambda\right)\right]\, ,\\
\label{c2.24} B^1&=&\frac{2\varphi_N}{c^2}-\frac{E^1}{a^2}=
\frac{2\varphi_N}{c^2}\left[1-\frac{1}{2}\exp\left(-r_3/\lambda\right)\right]\, .
\ea
%%%%%%%
It is well known that the metric correction term $A^1\sim O(1/c^2)$ describes the non-relativistic gravitational potential: $A^1=2\varphi/c^2$. Therefore, this potential
acquires the Yukawa correction term:
%%%%%%
\be{c2.25} \varphi = \varphi_N\left[1+\frac{1}{2}\exp\left(-r_3/\lambda\right)\right]\, . \ee
%%%%%%
The parameter $\lambda$ defines the characteristic range of Yukawa interaction. There is a strong restriction on this parameter from the inverse square law experiments.
For the Yukawa parameter $\alpha =1/2$ (which is the prefactor in front of the exponent) the upper limit is \cite{new}
%%%%%%
\be{c2.26}
\lambda_{max}=\left(\frac{a}{\sqrt{\omega_1}}\right)_{max}\approx6\times10^{-3}\ {\mbox{cm}}\, ,
\ee
%%%%%%
which provides the upper limit on the size $a$ of the internal two-sphere. The ratio $B^1/A^1$ goes to 1 in the limit $r_3\gg\lambda$. For example, for the gravitational
experiments in the Solar system (the deflection of light and the time delay of radar echoes) we can take $r_3\gtrsim r_{\odot}\sim 7\times 10^{10}$cm. Then, for $\lambda
\lesssim 6\times 10^{-3}$cm, we get $r_3/\lambda \gtrsim 10^{13}$. Therefore, with very high accuracy we can drop the Yukawa correction term, and the parameterized
post-Newtonian parameter $\gamma$ is equal to 1 similar to General Relativity, and we arrive at the concordance with the above-mentioned gravitational experiments.

Obviously, for $r_3\ll\lambda$ the ratio $B^1/A^1$ goes to 1/3. For the limiting case $\omega_1=0 \, \Rightarrow \, \lambda \to +\infty$, this ratio is exactly equal to
1/3. Therefore, the PPN parameter $\gamma =1/3$ which strongly contradicts the observations (see also \cite{ChEZ1} for the case of non-smeared extra dimensions). Exactly
the same result we have for the models with toroidal compactification of the internal spaces and a point-like gravitating mass with the dust-like equations of state in
the external and internal spaces (see \cite{EZ3,EZ4,EZ5} where $\gamma = 1/(D-2)$). The Eq. \rf{c2.7} shows that $\omega_1 =0$ for the models with $\Lambda_6=0$. On the
other hand, the parameter $\omega_1$ defines the Yukawa mass squared (see the Eq. \rf{c2.19}). Therefore, the positiveness of $\Lambda_6$ is the necessary condition for
the positiveness of the Yukawa mass squared in the models with spherical compactification. In appendix \ref{sec:C}, we show that this is also the sufficient condition of
the internal space stabilization.

%%%%%%%%%%%%%%%%%%%%%%%%%%%%%%%%%%%%%%%%%%%%%%%%%%%%%%%%%%%%%%%%%%%%%%%%%%%%%%%%%%%

%%%%%%%%%%%%%%%%%%%%%%%%%%%%%%%%%%%%%%%%%%%%%%%%%%%%%%%%%%%%%%%%%%%%%%%%%%%%%%%%%%%
\section{Conclusion and discussion}

To calculate the perihelion shift of planets and the deflection of light by the Sun, we need the metric coefficients in the weak-field limit. To perform the
corresponding calculations in General Relativity, we usually assume that the background space-time metrics is flat and perturbation has the form of a point-like mass
(see, e.g., \cite{Landau}). In our paper \cite{EZ3}, we generalized this procedure to the case of the extra dimensions. We considered flat background in the form of the
Kaluza-Klein model with toroidal compactification of the internal space, and perturbed this background by a point-like mass. We found that obtained formulas lead to a
strong contradiction with the observations. The exact soliton solutions considered in \cite{EZ4,EZ5} confirmed this result: the physically reasonable point-like massive
source contradicts the observations.  Among these solutions, latent solitons, in particular, black strings and black branes, are the only astrophysical objects which
satisfy the gravitational experiments at the same level of accuracy as General Relativity. However, their matter source does not correspond to a point-like mass with the
dust-like equations of state both in the external and internal spaces. In contrast, it has a very strange relativistic equation of state (tension) in the internal space.
Obviously, such equation of state requires careful physical justification. Up to now, we do not aware about it.

Further, trying to understand the underlying problem with a point-like massive source, we investigated non-linear $f(R)$ models with toroidal compactification of the
internal spaces \cite{EZnonlin}. Unfortunately, such modification of gravity does not save the situation. Here, to demonstrate the agreement with the observations at the
same level of accuracy as General Relativity, the gravitating massive source should also have tension in the internal spaces \cite{EZf(R)solitons}.

In the present paper, to avoid this problem, we considered the Kaluza-Klein model where the internal space is not flat but has the form of a two-sphere with the radius
$a$. Similar to General Relativity, the external space-time background remains flat and the perturbation takes the form of a point-like mass. Additionally, we included a
bare multidimensional cosmological constant and found that this is a crucial point for our model because it gives a possibility to stabilize the internal space. First,
we found the background matter which corresponds to our metric ansatz. It was shown that this matter can be simulated by a perfect fluid with the vacuum equation of
state in the external space and an arbitrary equation of state with the parameter $\omega_1$ in the internal space. Then, in the weak-field approximation we perturbed
the background matter and metrics by a point-like mass. We assumed that such perturbation does not change the equations of state. We have shown that in the case
$\omega_1>0$ the perturbed metric coefficients have the Yukawa type corrections with respect to the usual Newtonian gravitational potential. The inverse square law
experiments restrict such corrections and provide the following bound on the parameters of the model:  $a/\sqrt{\omega_1}\lesssim 6\times10^{-3}\ {\mbox{cm}}$.
Obviously, in the Solar system we can drop the Yukawa correction terms with very high accuracy, and the parameterized post-Newtonian parameter $\gamma$ is equal to 1
similar to General Relativity. Therefore, our model satisfies the gravitational experiments (the deflection of light and the time delay of radar echoes) at the same
level of accuracy as General Relativity. We have also found that our background matter provides the stable compactification of the internal space in the case
$\omega_1>0$. This is the main feature of our model with $\omega_1>0$. Neither the models with toroidal compactification in \cite{EZ3,EZ4,EZ5,EZnonlin} nor the model
with spherical compactification and $\omega_1=0$ (see also \cite{ChEZ1}) have this property. Therefore, we can achieve the agreement with the observations in models with
the stable compactification of the internal spaces. This is the main conclusion of our paper. The usual drawback of such models consists in fine tuning of their
parameters.

It is worth noting that the problem of stabilization of the internal spaces was extensively investigated in the literature in the framework of multidimensional
cosmological models including the Freund-Rubin and Casimir mechanisms (see, e.g., \cite{EZ4,exci,GKZ,GMZ}). Obviously, in this case we deal with the time-dependent
multidimensional metrics, where the four-dimensional part corresponds usually either to Friedmann or to DeSitter space-time. Our present paper is devoted to the
classical gravitational tests in the weak-field limit, e.g., in the Solar system, for Kaluza-Klein models. Clearly, this is an astrophysical problem with the static
gravitational field. Stabilization of the internal spaces in such astrophysical models was not investigated sufficiently in the literature. As far as we know, the
six-dimensional Kaluza-Klein model with spherical compactification of two extra dimensions is considered in detail with respect to observations in the Solar system for
the first time. We produce the consistent generalization of the weak-field approximation in General Relativity to the considered multidimensional case and obtain
solutions of Einstein equations in corresponding orders of approximation. Then, we explicitly demonstrate the crucial role of a bare six-dimensional cosmological
constant and restrict the parameters, proceeding from the experimental constraints. The background matter is taken in the form of the perfect fluid with initially
arbitrary equations of state. This choice is much more general than both Freund-Rubin two-forms and Casimir effect, which represent only particular cases. Our analysis
shows that without the multidimensional cosmological constant these two mechanisms can not provide the flat external space and at the same time the curved (and
stabilized!) internal space. Certainly, it is not evident from the very beginning. Moreover, our results enable us to estimate quantitatively the effect of extra
dimensions on gravitational tests and, vice versa, to put experimental limitations on the parameters of our multidimensional model.

%%%%%%%%%%%%%%%%%%%%%%%%%%%%%%%%%%%%%%%%%%%%%%%%%%%%%%%%%%%%%%%%%%%%%%%%%%%
\section*{ACKNOWLEDGEMENTS}
%We want to thank Prof. E.K. Loginov for useful discussions.
This work was supported in part by the "Cosmomicrophysics" programme of the Physics and Astronomy
Division of the National Academy of Sciences of Ukraine.

%%%%%%%%%%%%%%%%%%%%%%%%%%%%%%%%%%%%%%%%%%%%%%%%%%%%%%%%%%%%%%%%%%%%%%%%%%
%%%%%%%%%%%%%%
\appendix
\section{\label{sec:A} Components of the Ricci tensor}
\renewcommand{\theequation}{A\arabic{equation}}
\setcounter{equation}{0}

In this appendix we consider the six-dimensional space-time metrics of the form \rf{c2.9}:
%%%%%%
$$
ds^2=Ac^2dt^2+Bdx^2+Cdy^2+Ddz^2+Ed\xi^2+Fd\eta^2\, ,
$$
%%%%%%
where the metric coefficients $A,B,C,D,E$ and $F$ satisfy the decomposition \rf{c2.10}. Now, we define the corresponding components of the Ricci tensor up to linear
correction terms. In appendixes \ref{sec:A} and \ref{sec:B}, we assume that correction terms $A^1, B^1, C^1, D^1, E^1$ and $F^1$ may depend on all coordinates of the
five-dimensional space.

\subsection*{Diagonal components}
%%%%%
\be{a5} R_{00}\approx\frac{1}{2}\left[\triangle_3 A^1 +\frac{1}{a^2}\triangle_{\xi\eta}A^1\right]\, . \ee
%%%%%%%
Here we introduce the Laplace operators:
%%%%%%
\ba{a6} &{}& \triangle_3\equiv \frac{\partial^2}{\partial x^2} + \frac{\partial^2}{\partial y^2} + \frac{\partial^2}{\partial z^2}\, ,\nn\\
&{}& \triangle_{\xi\eta}\equiv \frac{\partial^2}{\partial \xi^2} + \frac{\cos \xi}{\sin \xi}\frac{\partial}{\partial\xi} + \frac{1}{\sin^2\xi}\frac{\partial^2}{\partial
\eta^2}\, . \ea
%%%%%
For $R_{11}$, $R_{22}$ and $R_{33}$ we obtain respectively:
%%%%%%%
\ba{a7} &{}& R_{11}\approx -\frac{1}{2}\left[-\triangle_3 B^1-\frac{1}{a^2}\triangle_{\xi\eta}B^1\right.\nn\\
&+&\left.\left(A^1+B^1-C^1-D^1-\frac{E^1}{a^2}-\frac{F^1}{a^2\sin^2\xi}\right)_{xx}\right]\, ,\nn\\
&{}& \ea
%%%%%%
%%%%%%
\ba{a8} &{}& R_{22}\approx -\frac{1}{2}\left[-\triangle_3 C^1-\frac{1}{a^2}\triangle_{\xi\eta}C^1\right.\nn\\
&+&\left.\left(A^1+C^1-B^1-D^1-\frac{E^1}{a^2}-\frac{F^1}{a^2\sin^2\xi}\right)_{yy}\right]\, ,\nn\\
&{}& \ea
%%%%%%
%%%%%%
\ba{a9} &{}& R_{33}\approx -\frac{1}{2}\left[-\triangle_3 D^1-\frac{1}{a^2}\triangle_{\xi\eta}D^1\right.\nn\\
&+&\left.\left(A^1+D^1-B^1-C^1-\frac{E^1}{a^2}-\frac{F^1}{a^2\sin^2\xi}\right)_{zz}\right]\, . \nn\\
&{}& \ea
%%%%%%%
Indexes denote everywhere the corresponding partial derivatives (e.g., $A_{x}\equiv \partial A/\partial x$). The components $R_{44}$ and $R_{55}$ read respectively:
%%%%%%%
\ba{a10} &{}& R_{44}\approx 1 -\frac {1}{2}
 \left\{(A^{1}-B^{1}-C^{1}-D^{1})_{\xi \xi} -\triangle_3 E^{1}\right.\nn\\
 &-&\left.\frac{E^{1}_{\eta \eta}}{a^2\sin^2\xi}-\frac{1}{a^2\sin^2\xi}\left[ \left( F^1_{\xi \xi}-
  2\frac{\cos2\xi}{\sin2\xi}F^1_\xi\right)\right.\right.\nn\\
 &-&\left.\left.\frac{2}{\sin 2\xi}\left( F^1_\xi - 2\frac{\cos\xi}{\sin\xi}F^1 \right)
  -\frac {\sin 2\xi}{2} E^1_{\xi} \right]
   \right\} \, ,
\ea
%%%%%
%%%%%
\ba{a11} &{}& R_{55}\approx \sin^2\xi-\frac {1}{2} \left\{(A^{1}-B^{1}-C^{1}-D^{1})_{\eta\eta}-\triangle_3 F^{1}\right.\nn\\
&-&\left.\frac{E^{1}_{\eta \eta}}{a^2} -\frac{1}{a^2}\left(  F^1_{\xi\xi}-2\frac{\cos2\xi}{\sin2\xi} F^1_{\xi} \right)\right.\nn\\
 &+&\left. \frac{\sin2\xi}{2} \left( \frac{E^1_\xi}{a^2}+\left(A^1-B^1-C^1-D^1\right)_\xi\right)\right.\nn\\
 &+&\left.\frac{\cos\xi}{a^2\sin\xi}\left( F^1_\xi -
2\frac{\cos\xi}{\sin\xi}F^1\right)\right.\nn\\
 &+&\left. \frac{1}{a^2}\frac{\sin\xi}{\cos\xi}
\left(  F^1_\xi - \sin{2\xi} E^1 \right)
\right\}\, .
\ea
%%%%%%%%%

\subsection*{Off-diagonal components}

%%%%%%%
Obviously, for the static metrics the components $R_{01}$, $R_{02}$, $R_{03}$, $R_{04}$, $R_{05}$ are identically equal to zero. Let us now calculate the remaining $10$
off-diagonal components:
%%%%%
\ba{a12}
 R_{12}\approx
\left(-\frac{1}{2}A^1+\frac{1}{2}D^1+\frac{1}{2a^2}E^1+\frac{1}{2a^2\sin^2\xi}F^1\right)_{xy}\, , \nn\\
\ea
%%%%%
%%%%%
\ba{a13}
 R_{13}\approx
\left(-\frac{1}{2}A^1+\frac{1}{2}C^1+\frac{1}{2a^2}E^1+\frac{1}{2a^2\sin^2\xi}F^1\right)_{xz}\, ,\nn\\
\ea
%%%%%%%
%%%%%%%
\ba{a14}
 R_{23}\approx
\left(-\frac{1}{2}A^1+\frac{1}{2}B^1+\frac{1}{2a^2}E^1+\frac{1}{2a^2\sin^2\xi}F^1\right)_{yz}\, ,\nn\\
\ea
%%%%%%%
%%%%%%%
\ba{a15}
 R_{15}\approx
\left(-\frac{1}{2}A^1+\frac{1}{2}C^1+\frac{1}{2}D^1+\frac{1}{2a^2}E^1\right)_{x\eta}\, , \ea
%%%%%%%
%%%%%%%
\ba{a16}
 R_{25}\approx
\left(-\frac{1}{2}A^1+\frac{1}{2}B^1+\frac{1}{2}D^1+\frac{1}{2a^2}E^1\right)_{y\eta}\, , \ea
%%%%%%
%%%%%%
\ba{a17}
 R_{35}\approx
\left(-\frac{1}{2}A^1+\frac{1}{2}B^1+\frac{1}{2}C^1+\frac{1}{2a^2}E^1\right)_{z\eta}\, , \ea
%%%%%%
%%%%%%
\ba{a18}
 &{}& R_{14}\approx \left(\frac{1}{2}( -A^1+C^1+D^1)_{\xi}+\frac{1}{2a^2\sin^2\xi}F_{\xi}^1\right.\nn\\
 &-& \left.\frac{\cos\xi}{2a^2\sin\xi}E^1-\frac{\cos\xi}{2a^2\sin^3\xi}F^1\right)_x\, ,\ea
%%%%%%
%%%%%%
\ba{a19}
 &{}& R_{24}\approx \left(\frac{1}{2}(-A^1+B^1+D^1)_{\xi}+\frac{1}{2a^2\sin^2\xi}F_{\xi}^1\right.\nn\\
 &-& \left.\frac{\cos\xi}{2a^2\sin\xi}E^1-\frac{\cos\xi}{2a^2\sin^3\xi}F^1\right)_y\, ,
\ea
%%%%%
%%%%%
\ba{a20}
 &{}& R_{34}\approx \left(\frac{1}{2}(-A^1+B^1+C^1)_{\xi}+\frac{1}{2a^2\sin^2\xi}F_{\xi}^1\right.\nn\\
 &-& \left.\frac{\cos\xi}{2a^2\sin\xi}E^1-\frac{\cos\xi}{2a^2\sin^3\xi}F^1\right)_z\, , \ea
%%%%%%
%%%%%%
\ba{a21}
 &{}& R_{45}\approx\left(\frac{1}{2}(-A^1+B^1+C^1+D^1)_{\xi}\right.\nn\\
 &+& \left.\frac{\cos\xi}{2\sin\xi}(A^1-B^1-C^1-D^1)\right)_{\eta}\, .
\ea
%%%%%%%

%%%%%%%%%%%%%%%%%%%%%%%%%%%%%%%%%%%%%%%%%%%%%%%%%%%%%%%%%%%%%%%%%%%%%%%%%%%%%%%%%%%%%%%%%%%%%%%%%%%%%%%%%%%%%%%%%%

\section{\label{sec:B} Relations between metric coefficients}
\renewcommand{\theequation}{B\arabic{equation}}
\setcounter{equation}{0}

First, we investigate expressions \rf{a12}-\rf{a14} in the case $R_{12}=R_{13}=R_{23}=0$. It can be easily seen that the equation $R_{12}=0$ has a solution
%%%
\ba{xxx} &{}&-\frac{1}{2}A^1+\frac{1}{2}D^1+\frac{1}{2a^2}E^1+\frac{1}{2a^2\sin^2\xi}F^1\nn\\
&=&C_1(z,\xi,\eta)f_1(x)+C_2(z,\xi,\eta)f_2(y)\, ,\nn\ea
%%%
where $C_1(z,\xi,\eta), C_2(z,\xi,\eta),
f_1(x)$ and $f_2(y)$ are arbitrary functions. We also assume that in the limit $|x|,|y|,|z|\to +\infty$ the perturbed metrics reduces to the background one. Thus, all
perturbations $A^1,B^1,C^1,D^1,E^1$ and $F^1$ as well as their partial derivatives vanish in this limit. Therefore, the right hand side of the above equation is equal to
zero. Similar reasoning can be applied to equations $R_{13}=0$ and $R_{23}=0$. Then, we arrive at the following relations:
%%%%%%%
\be{b1} B^1=C^1=D^1=A^1-\frac{1}{a^2}E^1-\frac{1}{a^2\sin^2\xi}F^1\, . \ee
%%%%%

We consider models where the Einstein equation for all off-diagonal components is reduced to
%%%%%
\be{b2}
R_{ik}=0 \quad \mbox{for}\quad i\neq k\, .
\ee
%%%%%
We want to analyze these equations for components \rf{a15}-\rf{a21} with regard to the relations \rf{b1}.

First, it can be easily seen that Einstein equations \rf{b2} for components \rf{a18}-\rf{a20} give
%%%%%%%
\ba{b3}
 &{}& -A_{\xi}^1+B_{\xi}^1+C_{\xi}^1+\frac{1}{a^2\sin^2\xi}F_{\xi}^1-\frac{\cos\xi}{a^2\sin\xi}E^1\nn\\
 &-& \frac{\cos\xi}{a^2\sin^3\xi}F^1=C_3(\xi,\eta)\, , \ea
%%%%%
where $C_3(\xi,\eta)$ is an arbitrary function. From the boundary conditions at $|x|,|y|,|z|\to +\infty$ we find that $C_3(\xi,\eta)=0$. Taking it into account, we get
from \rf{b1} and \rf{b3} respectively
%%%%%
\be{b4}
-A^1+B^1+\frac{1}{a^2\sin^2\xi}F^1=-\frac{1}{a^2}E^1
\ee
%%%%%%
and
%%%%%%
\be{b5}
-A_{\xi}^1+B_{\xi}^1+\frac{1}{a^2\sin^2\xi}F_{\xi}^1-\frac{\cos\xi}{a^2\sin\xi}E^1-\frac{\cos\xi}{a^2\sin^3\xi}F^1=-B_{\xi}^1\, .
\ee
%%%%%
Differentiating \rf{b4} with respect to $\xi$, we obtain
%%%%%%
\be{b6}
-A^1_{\xi}+B^1_{\xi}+\frac{1}{a^2\sin^2\xi}F^1_{\xi}-\frac{2\cos\xi}{a^2\sin^3\xi}F^1=-\frac{1}{a^2}E^1_{\xi}\, .
\ee
%%%%
Subtraction \rf{b6} from \rf{b5} yields
%%%%%
\be{b7}
\frac{1}{a^2\sin^2\xi}F^1 =-B_{\xi}^1\tan\xi+\frac{1}{a^2}E^1_{\xi}\tan\xi+\frac{1}{a^2}E^1\, .
\ee
%%%%%

Let us consider now the case of the smeared extra dimensions.

\vspace{0.3cm}

{\it{ Smeared extra dimensions}}

\vspace{0.3cm}

In this case, the matter source is uniformly smeared over the internal space \cite{footnote 3}. It results in the metric coefficients $A^1,B^1,C^1,D^1$ and $E^1$
depending only on the external coordinates $x,y$ and $z$ \cite{EZ2}. We require that  the off-diagonal components must be like \rf{b2}. Then, equations
$R_{15}=R_{25}=R_{35}=R_{45}=0$ (where these off-diagonal components are defined by \rf{a15}-\rf{a17},\rf{a21}) are automatically satisfied. It can be easily seen from
\rf{b1} that the coefficient $F^1 \sim \sin^2\xi$. Moreover, to satisfy the Eq. \rf{b7}, it should have the form
%%%%%%
\be{b8}
F^1=E^1\sin^2\xi\, .
\ee
%%%%%%
Therefore, \rf{b1} can be rewritten in the form
%%%%%
\be{b9}
-A^1 + B^1 +\frac{2}{a^2}E^1=0\, .
\ee
%%%%%

%%%%%%%%%%%%%%%%%%%%%%%%%%%%%%%%%%%%%%%%%%%%%%%%%%%%%%%%%%%%%%%%%%%%%%%%%%
%%%%%%%%%%%%%%

\section{\label{sec:C}Stabilization of the internal two-sphere}
\renewcommand{\theequation}{C\,\arabic{equation}}
\setcounter{equation}{0}

To consider the stabilization of the internal space, we suppose that the scale factor of the internal space becomes a function of time: $a \to a(t)$. Then, the energy
conservation equation has the simple integral for the energy density (see, e.g., (2.9) in \cite{Zhuk})
%%%%%
\be{6} T^i_{k;i}=0\; \quad \Rightarrow \quad \varepsilon (t)=\frac{\tilde \varepsilon_{c}}{(a(t))^{2(\omega_1+1)}}\, , \ee
%%%%%%
where $\tilde \varepsilon_{c}$ is a constant of integration. Let us introduce the following notation:
%%%%%
\be{7}
a(t)\equiv e^{\beta (t)}=e^{\beta_0+\tilde\beta(t)}=a\, e^{\tilde \beta(t)}\, ,
\ee
%%%%%
where $a=\exp(\beta_0)=\mbox{const}$ is some initial value or a position of the stable compactification. The latter corresponds to a minimum of an effective potential
(see below). Hence, the Eq. \rf{6} for the energy density reads
%%%%%%
\be{8} \varepsilon (t) = \varepsilon_{c} e^{-2(\omega_1+1)\tilde\beta(t)}\, ,\quad \varepsilon_{c}\equiv  \frac{\tilde \varepsilon_{c}}{a^{2(\omega_1+1)}}=\mbox{const}\,
. \ee
%%%%%%
The mechanism of the internal space stabilization was described in detail in \cite{Zhuk}. To get such stabilization, we should find a minimum (which corresponds to
$\tilde \beta =0$) of the effective potential
%%%%%%%
\be{9}
U_{eff}(\tilde \beta)= e^{-2\tilde\beta}\left[-\frac12 \tilde R_1 e^{-2\tilde \beta}+\kappa\Lambda_6 +\kappa  \varepsilon_{c} e^{-2(\omega_1+1)\tilde \beta}\right]\, ,
\ee
%%%%%%%
where $\tilde R_1\equiv 2/a^2 =\mbox{const}$ is the scalar curvature of the internal space (two-sphere). The minimum of this potential defines an effective
four-dimensional cosmological constant: $\Lambda_{(4)eff}=U_{eff}(\tilde\beta=0)$. We consider the case of flat external space-time. Therefore, the effective
cosmological constant should be equal to zero:
%%%%%
\be{10}
\Lambda_{(4)eff}=0\, \quad \Rightarrow \quad \frac12 \tilde R_1 =\kappa\Lambda_6+\kappa \varepsilon_{c}\, .
\ee
%%%%%%
The comparison of this equation with the first one in \rf{c2.5} shows that $\bar \varepsilon \equiv \varepsilon_{c}$. Therefore, the Eq. \rf{10} is automatically
satisfied for our model.

The extremum condition gives
%%%%%%
\be{11} \left.\frac{\partial U_{eff}}{\partial\tilde \beta}\right|_{\tilde \beta=0}=0\, \quad \Rightarrow \quad \frac12 \tilde R_1=(\omega_1+1)\kappa  \varepsilon_{c}\,
. \ee
%%%%%%
The positiveness of $\tilde R_1$ and $\varepsilon_{c}$ results in the condition $\omega_1>-1$. From Eqs. \rf{10} and \rf{11} we get the fine tuning
%%%%%%
\be{12}
\Lambda_6=\omega_1 \varepsilon_{c}\, .
\ee
%%%%%%
It can be easily verified that this condition follows also from Eqs. \rf{c2.5} and \rf{c2.7}. Therefore, it is automatically satisfied for our model. To get the
stabilization, the extremum should be a minimum:
%%%%%%
\be{13}
\left.\frac{\partial^2 U_{eff}}{\partial\tilde \beta^2}\right|_{\tilde \beta=0}= 4\omega_1(\omega_1+1)\kappa \varepsilon_{c}>0\, ,
\ee
%%%%%
where we have used the condition \rf{10}. It can be easily seen that the minimum takes place for $\omega_1>0$ which exactly coincides with the condition \rf{c2.21}.
Therefore, the case of dust $\omega_1=0$ (the \cite{ChEZ1} case) does not fit this condition because the effective potential is flat.

For example, in the case of the Freund-Rubin stable compactification (see, e.g., \cite{FR,exci,Zhuk}) with two-forms
%%%%%%
\be{14}
F_{ik}=\left\{
\begin{array}{cc}
\sqrt{|g_2|}\, \varepsilon_{ik} f & \mbox{for   }\,  i,k=4,5\\
\\
0 & \mbox {otherwise}\,
\end{array} \right . \quad
\ee
%%%%%%
where $|g_2|=|g_{44}g_{55}| =a^4\sin^2\xi$ is the determinant of the metrics on the sphere of the radius $a$, and $\varepsilon_{ik}$ is a totally antisymmetric
Levi-Civita tensor with $\varepsilon_{45}=-\varepsilon_{54}=1$, and $f$ is a constant which we define below, the energy-momentum tensor is
%%%%%%%
\be{15}
T_{ik}=\left\{
\begin{array}{cc}
\left(f^2/(8\pi)\right) g_{ik} & \mbox{for   }\,  i,k=0,...,3\, ;\\
\\
-\left(f^2/(8\pi)\right) g_{ik} & \mbox{for   }\,  i,k=4,5\, .
\end{array} \right . \quad
\end{equation}
%%%%%%%
The comparison of this expression with the background energy-momentum tensor \rf{c2.3} shows that the parameter of the equation of state in the internal space
$\omega_1=1$. Then, it can be easily seen with the help of Eqs. \rf{c2.7}, \rf{8} and \rf{12} that we get the following fine tuning relations:
%%%%%
\be{16}
\frac{f^2}{8\pi}=\Lambda_6 = \frac{1}{2\kappa a^2}=\bar \varepsilon= \tilde \varepsilon_{c}/a^{4}\,
\ee
%%%%%%
with full agreement with the results of the paper \cite{Zhuk}. Similarly, we can consider the stabilization by means of the Casimir effect where $\omega_1 = 2$
\cite{exci}.

%%%%%%%%%%%%%%%%%%%%%%%%%%%%%%%%%%%%%%%%%%%%%%%%%%%%%%%%%%%%%%%%%%%%%%%%%%%%%%%%


\begin{thebibliography}{99}
\bibitem{footnote}
Such equations of state take place in the external and internal spaces for a point-like mass at rest. For ordinary astrophysical extended objects, e.g., for the Sun, it
is also usually assumed that the energy density is much greater than the pressure.
\bibitem{footnote 2}
The case of an arbitrary matter source was considered in \cite{ChEZ1} for the model with $\Lambda_6=0$.
\end{thebibliography}
\end{document}